\documentclass[12pt]{article}

\usepackage{amsmath,amssymb,amsthm,epsf,epsfig}
\usepackage{amsmath,amssymb}

\begin{document}
\begin{center}
\large{\textbf{\textbf{SPECTRAL DISCONTINUITIES IN CONSTRAINED DYNAMICAL MODELS}}}\\
\end{center}
\begin{center}
Sudipta Das {\footnote{E-mail: sudipta\_jumaths@yahoo.co.in} and
Subir Ghosh {\footnote{E-mail: sghosh@isical.ac.in}\\
Physics and Applied Mathematics Unit, Indian Statistical
Institute\\ 203 B.T.Road, Kolkata 700108, India}}
\end{center}

\vskip 1cm {\it{\textbf{Abstract}}}:
As examples of models having interesting constraint structures, we derive a
quantum mechanical model from the spatial freezing of a well known
relativistic field theory - the chiral Schwinger model. We apply the
Hamiltonian constraint analysis of Dirac \cite{dir} and find that the
nature of constraints depends critically on a $c$-number parameter
present in the model. Thus a change in the parameter  alters the
number of dynamical modes in an abrupt and non-perturbative way.

We have obtained new  {\it{real}} energy levels for the quantum
mechanical model  as  we explore {\it{complex}} domains in the
parameter space. These  were forbidden in the parent chiral
Schwinger field theory where the analogue Jackiw-Rajaraman
parameter is restricted to be real. We explicitly show existence
of modes that satisfy higher derivative Pais-Uhlenbeck form of
dynamics \cite{Uh}.

We also show that the Cranking Model \cite{val}, well known in Nuclear Physics,
can be interpreted as a spatially frozen version of another well studied
relativistic field theory in $2+1$-dimension- the
Maxwell-Chern-Simons-Proca Model \cite{djt}.

\section{Introduction:} The Hamiltonian formulation of constrained
dynamical systems, as formulated by Dirac \cite{dir}, provides a systematic
framework to analyze and
quantize constrained systems. In this scheme there is an extremely
important classification of constraints: First Class Constraints
(FCC) and Second Class Constraints (SCC) (for a brief discussion
see Section 2) and these two types of constraints act in qualitatively
different manners. In a particular model with constraints it might
happen that as one moves smoothly in the parameter space the
nature of the constraint system changes from FCC to SCC. Clearly
this will lead to a dramatic (and abrupt) change in the  spectra
and dynamics even though the model (Lagrangian or Hamiltonian) itself will not
show any drastic change (such as the appearance of an explicit
singularity or otherwise). What we mean by this behavior will become clear
later when we discuss a
specific model. This is quite contrary to the normal behavior of a
system towards a change in its parameters where small changes in
the parameter is reflected in a small change in the dynamics.
{\it{The reason, in the present case, is that the change in the
parameter value is associated with a change in the constraint
structure governing the system.}} The change in the nature of the
constraints induces a non-perturbative change in the entire
system.  In physical terms, as the constraint structure changes
from SC to FC, the system gains more symmetry (in the form of
local gauge invariance) and this is reflected in a more restricted
dynamics  in the FCC system than the SCC system. This phenomenon
can be observed in a generic model where a physical and dynamical
mode present in the SCC system  abruptly vanishes at that
particular point in parameter space where the system becomes FCC.
This is nicely revealed in the specific example we provide in this
paper.

In the present article, we will demonstrate that both in the particle model and
its parent field theoretic model, an entire Harmonic Oscillator mode
(in the former) and its relativistic field theoretic analogue - a
massive Klein-Gordon mode (in the latter) disappears as one passes
from the SCC to the FCC system. We stress that this passage in the
parameter space is smooth as far as the explicit expressions of
Lagrangian or Hamiltonian of the model is concerned and the
non-perturbative changes in spectra and dynamics are seen only after a proper
constraint analysis of the systems.

Moreover in the particle model, we will also study the model for
complex values of the parameter and show that it can yield
physical (real) energy states which reminds us of similar behavior
in non-Hermitian $PT$-symmetric models \cite{Bender}. In the field
theoretic example we will show that at the crossover point in
parameter space the dynamics is governed by a higher order
dynamical equation which can be thought of as a field theory
analogue of the Pais-Uhlenbeck oscillator \cite{Uh} of revived
interest \cite{pais}.

The paper is structured as follows: In Section 2 we give a brief
discussion on constraints that is relevant for the present paper.
In Section 3 we study the model that has constraints and the
constraint classification depends on a parameter value present in
the model. We analyze this singular point in parameter space in
detail. Next we move on to complex values of the parameter and
obtain some hitherto unknown results for this particular model.
This finite dimensional model has been derived from a very well
known field theory model, the bosonized chiral Schwinger model
\cite{sch,jr}, that we study briefly in Section 4. In fact the
connection with the Pais-Uhlenbeck model will be made in this
section. In Section 5 we provide another example of a similar identification
between well studied finite dimensional and field theory models:
the Cranking model \cite{val} and Maxwell-Chern-Simons-Proca field
theory \cite{djt,rb,sg}. We end with concluding remarks in Section
6.

\section {\bf {Constraint Analysis:}} In the Hamiltonian formulation
of constrained system \cite{dir} any relation between dynamical
variables, not involving time derivative is considered as a
constraint. Constraints can appear from the construction of the
canonically conjugate momenta (known as Primary constraint) or
they can appear from demanding time invariance of the constraints
(Secondary constraint).

Once the full set of constraints is in hand they are classified as
FCC or SCC according to whether the constraint Poisson bracket
algebra is closed or not, respectively. Presence of constraints
indicate a redundance of Degrees Of Freedom (DOF) so that not all
the DOFs are independent. FCCs signal presence of local gauge
invariances in the system. If FCCs are present, there are two ways
of dealing with them. Either one keeps all the DOFs but imposes
the FCCs by restricting the set of physical states to those
satisfying $(FCC)\mid state >=0$. On the other hand one is allowed
to choose further constraints, known as gauge fixing conditions so
that these together with the FCCs turn in to an SCC set and these
will also give rise to Dirac brackets that we presently discuss.
In case of say two SCCs, say for $(SCC)_1,(SCC)_2$ with the Poisson bracket
$\{(SCC)_1,(SCC)_2\}={\cal{C}}$
where ${\cal{C}}\ne 0$ is not another constraint, proceeding as before with
$(SCC)\mid state >=0$ one reaches an inconsistency because in the
identity $<state \mid\{(SCC)_1,(SCC)_2\}\mid state >=<state \mid {\cal{C}} \mid
state>$ the $LHS=0$ but $RHS\neq 0$. For consistent imposition of
the SCCs one defines the Dirac brackets between two generic
variables $A$ and $B$,
\begin{equation}
\{A,B\}_{DB}=
\{A,B\}-\{A,(SCC)_i\}\{(SCC)_i,(SCC)_j\}^{-1}\{(SCC)_j,B\},\label{di}
\end{equation}
where $(SCC)_i$ is a set of SCC and $\{(SCC)_i,(SCC)_j\}$ is the
constraint matrix. For SCCs this matrix is invertible (for finite
dimensional bosonic system, the number of SCCs is always even) and
since $\{A,SCC_i\}_{DB}=\{SCC_i,A\}_{DB}=0$ for all $A$ one can
implement $SCC_i=0$ strongly meaning that some of the DOFs can be
removed thereby reducing the number of DOFs in the system but one
must use the Dirac brackets in all subsequent computations. Hence,
SCCs induce a change in the symplectic structure and subsequently
one quantizes the Dirac brackets. Same principle is valid for
gauge fixed FCC system mentioned before. Hence, to understand the
effect of constraints we note that the presence of one FCC
(together with its gauge fixing constraint) or SCC can remove two
or one DOF from phase space respectively. We will apply this
scheme in a specific model.

\section {\bf{Particle Model:}} Let us consider the following
Lagrangian,
\begin{equation}
L = \frac{1}{2} \dot{A_{1}}^{2} + \frac{1}{2} \dot{\phi}^{2} + e
(A_{0} \dot{\phi} +  \dot{A_{1}} \phi )+ \frac{a
e^{2}}{2}(A_{0}^{2} - A_{1}^{2})\label{n1}
\end{equation}
where an overdot represents the derivatives with respect to time
and $a$ and $e$ are numerical parameters. $A_0,A_1,\phi$
constitute the dynamical variables. The somewhat unconventional
notation will become clear when we connect this model with the
field theory \cite{jr}, where $A_i, i=0,1$ and $\phi $ will become
electromagnetic gauge potentials and a scalar field respectively
with $e$ being the electric charge. Hence, although not mandatory,
we prefer to keep $e$ unchanged and explore the parameter space by
varying $a$. \\
 The conjugate momenta are $ \pi = \frac{\partial L}{\partial
\dot{\phi}}=\dot{\phi}+e A_{0}~~$,$~~\pi^{1}=\frac{\partial
L}{\partial \dot{A_{1}}}=\dot{A_{1}}+e
\phi~~$,$~~\pi^{0}=\frac{\partial L}{\partial \dot{A_{0}}}=0$ and
one immediately notices a Primary constraint
\begin{equation}
\psi_{1} \equiv \pi_{0} \approx 0.  \label{c1}
\end{equation}
The Hamiltonian is computed as,
\begin{equation}
H = \pi \dot{\phi} + \pi^{0} \dot{A_{0}} + \pi^{1} \dot{A_{1}} - L
= \frac{1}{2}\left[(\pi_{1} + e \phi)^{2} + (\pi - e A_{0})^{2} -
a e^{2}(A_{0}^{2} - A_{1}^{2})\right] +\lambda \pi _0, \label{h1}
\end{equation}
where we append the constraint $\psi_1$ through a Lagrange
multiplier. The canonical Poisson brackets are
$$\{A_{\mu},\pi^{\nu}\}= g_{\mu}^{\nu}~,~\{\phi ,\pi \}=1.$$ We
use the metric $g_{\mu \nu}=diag(1 , -1)$ and $\epsilon^{01}=1$.

Time persistence of the primary constraint $\psi_{1}$  leads to
the secondary constraint,
\begin{equation}
\psi_{2} \equiv \dot{\psi_{1}} = \{\pi_0,H\}= \pi + e (a-1) A_{0}
\approx 0. \label{c2}\end{equation} Note that there are no further
constraints since $\dot \psi_2$ will not yield a new constraint
but only fix the Lagrange multiplier $\lambda $.  This happens because
the pair $\psi_i,\psi_2 $ are SCC as we find out below.

The non-vanishing constraint bracket,
\begin{equation}
\{\psi_1,\psi_2\}=-e(a-1), \label{c3}
\end{equation}
shows that for $a \ne 1$ the set $\psi _i$ is SCC. Clearly this is
an explicit example of the interesting scenario that we mentioned
earlier because for $a=1$ this set is not SCC and in fact we will later
show that actually $\pi _0 =0$ turns out to be an FCC for the
special case $a=1$. However, clearly there is no significant
qualitative change or singular behavior in the {\it{explicit
expression}} of the Lagrangian (\ref{n1}) or the
Hamiltonian (\ref{h1}) as we pass through the point $a=1$ in the
parameter space for positive $a$.\\

$\mathbf{a\ne 1}:$~~ We now stick to $a\ne 1$ and compute the
constraint matrix along with its inverse,
\begin{equation}
\{ \psi_i,\psi_j\}=
 \left (
\begin{array}{cc}
 0 &  -e(a-1)\\
e(a-1) &  0
\end{array}
\right ) \label{mat}~~~~,~~~~
\{ \psi_i,\psi_j\}^{-1}=
 \left (
\begin{array}{cc}
 0 &  \frac{1}{e(a-1)}\\
-\frac{1}{e(a-1)} &  0
\end{array}
\right )
\end{equation}

Use of
(\ref{di}) leads to the Dirac brackets
\begin{equation}
 \{A_{0} , \phi\}_{DB} =
\frac{1}{e(a-1)},~\{A_{1} , \pi_{1}\}_{DB} = -1, ~\{\phi ,
\pi\}_{DB} = 1, \label{c4}
\end{equation}
 and the reduced Hamiltonian,
\begin{equation}
H = \frac{1}{2}\left[(\pi_{1} + e \phi)^{2} + a e^{2} A_{1}^{2} +
\frac{a}{a - 1} \pi^{2}\right].
\label{cc4}
\end{equation}
Notice that the singular behavior at $a\rightarrow 1$ is now manifest in the
Dirac brackets (\ref{c4}) or the reduced Hamiltonian (\ref{cc4}) but it
is only after we have properly taken care of the constraints.

However, it is crucial to note that the $a=1$ point has to be considered
separately since the constraints $\psi_i$s
{\it{commute}}, i.e. $\{\psi_{1},\psi_{2}\}=0$ and hence are FCC
and the system has to be dealt in a completely different way to
which we will come to later. But for now suffice it to say that
$a=1$ has a special significance.

From the following equations of motion $ \dot{\phi} = \{\phi ,
H\}_{DB} = \frac{a}{a - 1} \pi$~,~$\dot{\pi} =
 - e(\pi_{1} + e \phi)$~,~$\dot{A_{1}}  = -(\pi_{1}
+ e \phi)$~,~$\dot{\pi_{1}}  =  a e^{2} A_{1}$ we recover the
spectra:
\begin{equation}
({\pi_{1}} + e {\phi})^{..} = -\frac{a^{2} e^{2}}{a -
1}\left(\pi_{1} + e \phi\right);~~ ({\pi} - e {A_{1}})^{..} =
0.\label{s1}
\end{equation}
From the bracket $\{(\pi - e A_{1}) , (\pi_{1} + e \phi)\}_{DB}=
0$ we find that the model is, in fact, free. For convenience we
rename the variables:~~$
A_{1}=x_{1}$~,~$\phi=x_{2}$~,~$\pi_{1}=p_{1} - e \phi=p_{1}-e
x_{2}$~,~$\pi=p_{2}+e A_{1}=p_{2}+e x_{1}$ such that
$(x_{1},p_{1})$ and $(x_{2},p_{2})$ constitute two independent
canonical pairs. The Hamiltonian and dynamical equations are,
\begin{equation}
H = \frac{1}{2}\left[p_{1}^{2} + p_{2}^{2} + \frac{a^{2} e^{2}}{a
- 1}\left(x_{1} + \frac{p_{2}}{a e}\right)^{2}\right],
\end{equation}
\begin{equation}
\dot{p_{2}} = 0~~;~~\ddot{p_{1}} = - \frac{a^{2} e^{2}}{a - 1}
p_{1} \equiv -\omega^{2}p_{1}~~;~~\ddot{x_{1}} = -
\omega^2 x_{1}.\label{p1}
\end{equation}
Clearly we are dealing with a Harmonic Oscillator (HO) and a
decoupled free particle. Since $p_2$ is a constant with suitable
boundary condition we put $p_2=0$ and end up with $H =
\frac{1}{2}\left[p_{1}^{2} + \frac{a^{2} e^{2}}{a - 1}
x_{1}^{2}\right]$. Hence we have an HO with frequency $\omega $
satisfying $ \omega^{2} = \frac{a^{2} e^{2}}{a - 1}$.

Notice that the HO frequency, or equivalently the quantized HO
energy becomes complex for $a<1$ and is real for $a>1$. Apparently
the energy diverges for $a=1$ signalling a singularity. However,
as we have already mentioned, we have to treat the $a=1$ case
separately since the constraint structure shifts from SCC to FCC
for $a=1$. We show that the theory is indeed regular at $a=1$ but
with a different spectra.\\

$\mathbf{a =1}$: ~~Let us return to the starting Lagrangian
(\ref{n1}) and put directly $a=1$,
\begin{equation}
L = \frac{1}{2}\dot{A_{1}}^{2} + \frac{1}{2} \dot{\phi}^{2} +
eA_{0} \dot{\phi}+e\phi \dot{A_{1}}+ \frac{1}{2} e^{2} (A_{0}^{2}
- A_{1}^{2})\label{p5}.
\end{equation}
Similar analysis as before now yields the momenta, $ \pi =
\dot{\phi} + e A_{0} $~,~$\pi^{0} =0 $~,~$\pi^{1} = \dot{A_{1}}+e
\phi$ and the Hamiltonian:
\begin{equation}
H=\frac{1}{2} \pi^{2} +\frac{1}{2}e^{2} \phi^{2} +
\frac{1}{2}e^{2} A_{1}^{2} + \frac{1}{2} \pi_{1}^{2} + e \phi
\pi_{1}- e \pi A_{0}.
\end{equation}
Now we have three constraints:~~$ \chi \equiv \pi_{0} \approx
0$~,~$ \psi_{1} \equiv \dot{\chi} = \{\chi,H\} = \pi \approx
0$~,~$\psi_{2} \equiv \dot{\psi_{1}} = \{\psi_{1},H\} = \pi_{1}+e
\phi \approx 0$, \label{a1} where $ \chi $ is an FCC since it
commutes with all constraints, i.e. $\{\chi,\psi_{i}\}=0$ and
$\psi_i$ are SCC since $\{\psi_1,\psi_2\}=-e$. Presence of the FCC
$\chi $ allows us to choose another constraint $G\equiv A_0=0$ as
a gauge condition such that the set $\{G,\chi\}=1$ becomes a set
of SCC. Hence in this particular gauge we have a completely SCC
system of four constraints. Now the Dirac brackets and the reduced
Hamiltonian $H$
are~~$\{A_{1},\phi\}_{DB}=\frac{1}{e}$~,~$\{A_{1},\pi_{1}\}_{DB}=-1$~;~~~$
H = \frac{1}{2} e^{2} A_{1}^{2}$. Re-scaling the
variables,~~$-eA_1 \equiv p~,~\frac{1}{e}\pi_{1} \equiv x$ we find
just the trivial dynamics of a free particle:
\begin{equation}
H=\frac{p^2}{2}~~;~~\ddot{x} = 0\label{es}.
\end{equation}
Let us pause to observe that the spectra (\ref{es}) (for $a = 1$)
is very different from the previous case (\ref{p1}) (for $a\ne
1$). We find that the HO excitation is absent for $a=1$ due to
additional (gauge) symmetry in the system that further restricts
the dynamical content. Physically as $a\rightarrow 1$ the HO
frequency diverges so that its motion averages out and the
excitation drops out from the spectrum. But this amounts to a form
of coalescence of energy levels because as one approaches $a=1$
from $a<1$ (see Figure 1) it appears that the imaginary part of
the frequencies $\omega _{\pm}=\pm i\frac{ae}{\sqrt{a-1}} $ will
end up at $\mp\infty $ respectively (and hence will not meet) but
that is not the case since at $a=1$ the HO energy is effectively
zero. Hence the full tower of HO states coalesce to the ground
state (we will comment on this at the end).\\
In all the figures we have taken $e=1$.
\begin{figure}
\includegraphics[height=2.5in]{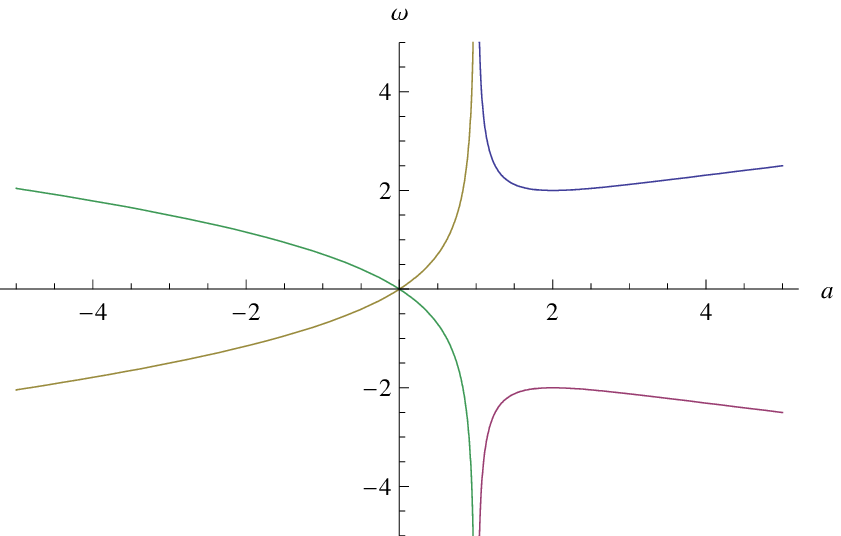}\\
\vspace{1mm}~~~~~~~~~~~~~~~~~~~~~~~~~~~~~~~~~~~~~~~~Fig.1\\
{\it{Fig. 1 shows the variation of $\omega$ against $a$ ($a$ is
real). The blue and red lines correspond to the positive real part
and negative real part of $\omega$ respectively; the green and
yellow lines correspond to the positive imaginary part and
negative imaginary part of $\omega$ respectively. Here the energy
$\omega$ has bound $|Re[\omega]| \geq 2$.}} \vspace{1mm}
\end{figure}\\

{\it {\textbf{Complex $a$}}}:~~ As we mention in Section 4, in the field
theoretic model \cite{jr} $a$ is a real number but for the present
particle model we are free to consider complex $a=a_0 +ia_1$ and
{\it{a priori}} we have a non-hermitian model in (\ref{n1}).  The
explicit form of energy for complex $a$ is,
\begin{equation}
\omega=\pm~e\left((\sqrt{(a_{0}-1)^{2}+a_{1}^{2}}+1)
\frac{(\sqrt{(a_{0}-1)^{2}
+a_{1}^{2}}+a_{0}-1)^{\frac{1}{2}}}{(2((a_{0}-1)^{2}
+a_{1}^{2}))^{\frac{1}{2}}}
\right.$$$$\left.+i(\sqrt{(a_{0}-1)^{2}+a_{1}^{2}}-1)
\frac{(\sqrt{(a_{0}-1)^{2}
+a_{1}^{2}}-a_{0}+1)^{\frac{1}{2}}}{(2((a_{0}-1)^{2}
+a_{1}^{2}))^{\frac{1}{2}}}\right)\label{n3}.
\end{equation}
For $a_1=0$ it reduces to the previous one (\ref{p1}). However, we
find new results by requiring $\omega $ to be real even with
complex $a$ ($a_1\ne0$). We find two interesting
consequences:~~(i) There is a non-trivial relation giving rise to
the separate bounds:
\begin{equation}
(a_{0} - 1)^{2} + a_{1}^{2} = 1~~;~~\mid a_1 \mid <1~,~\mid a_0 -1
\mid <1 \label{n2}.
\end{equation}
Using (\ref{n2}) we find that $\omega$ reduces to real values
$\omega =\pm e\sqrt {2a_0}$ and {\it{ the singularity of $\omega $
disappears}}. Furthermore, we can have real and positive energies
even for $\mid a \mid =\sqrt{(a_0)^2+(a_1)^2}<1$. This can be
contrasted with real $a$ where only $a>1$ will yield positive and
real energy as in (\ref{p1}).\ (ii) For $a_1\ne 0$, due to the
relation (\ref{n2}), a restriction is imposed on the real part
$a_0$. Again for real $a$ there appears no such restriction apart
from $a>1$ as in (\ref{p1}).\\

{\it{\textbf{Discussion of the Figures 2-5}}}:
\begin{figure}
\includegraphics[height=2in]{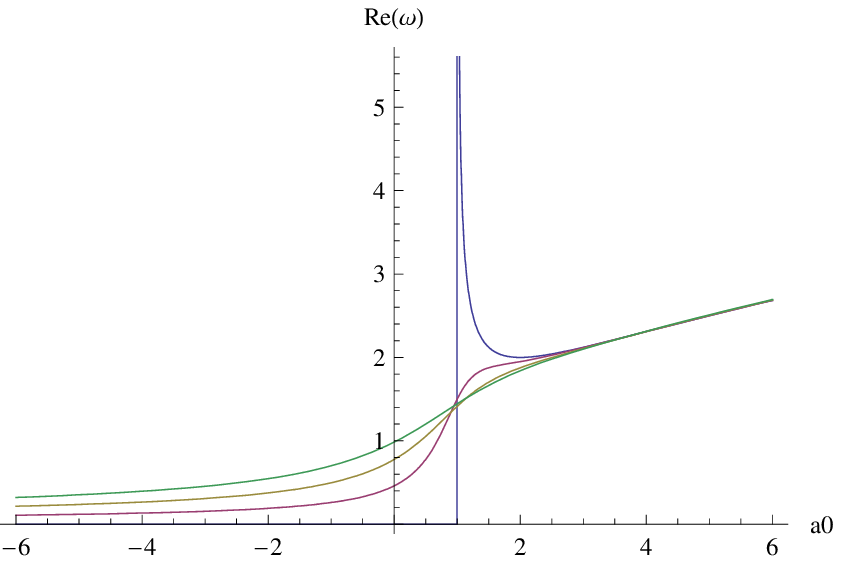}~
\includegraphics[height=2in]{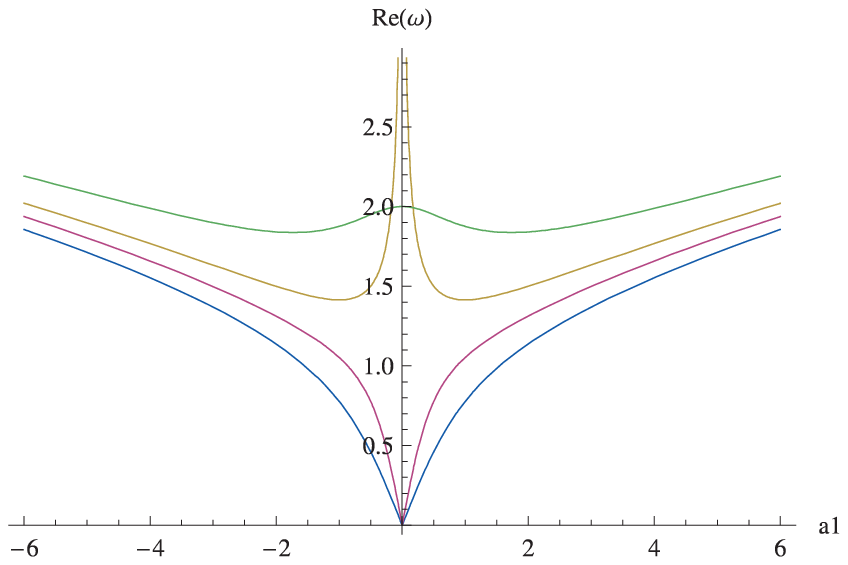}\\
\vspace{1mm} ~~~~~~~~~~~~~~~~~~~~~~~Fig.2~~~~~~~~~~~~~~~~~~~~~~~
~~~~~~~~~~~~~~~~~~~~~~~~~~~~Fig.3\\

{\it{Figs. 2 and 3 reproduce the variation of positive
value of $Re[\omega]$ against $a_{0}$ (keeping $a_{1}$ fixed) and
against $a_{1}$ (keeping $a_{0}$ fixed) respectively. In Fig. 2,
the different values of $a_{1}$ are: 0(blue), .5(red), 1(yellow),
1.5(green) and in Fig. 3, the different values of $a_{0}$ are:
0(blue), .5(red), 1(yellow), 2(green)}}. \vspace{1mm}
\end{figure}

\begin{figure}
\includegraphics[height=2in]{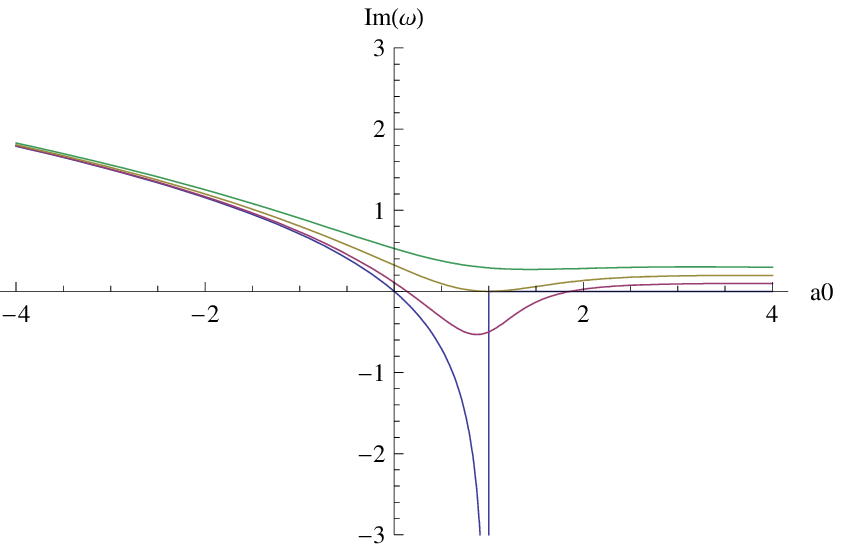}~
\includegraphics[height=2in]{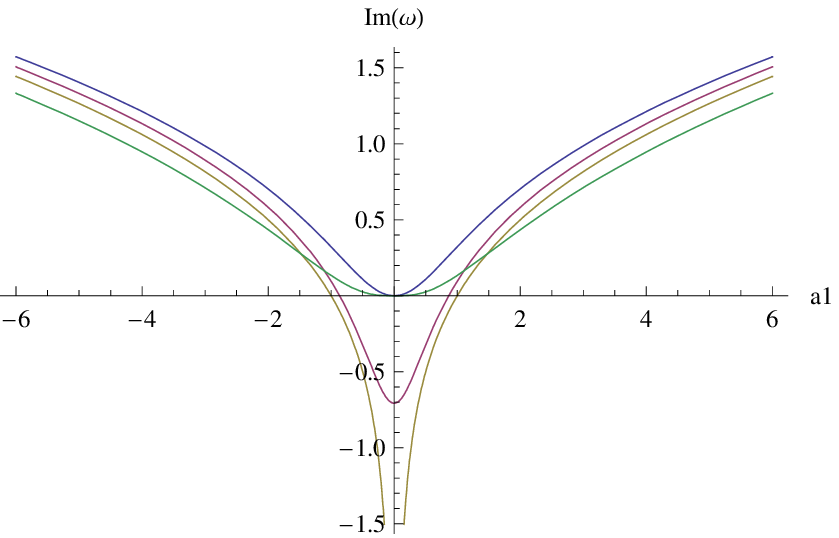}\\
\vspace{1mm} ~~~~~~~~~~~~~~~~~~~~~~~Fig.4~~~~~~~~~~~~~~~~~~~~~~~
~~~~~~~~~~~~~~~~~~~~~~~~~~~~Fig.5\\

{\it{Figs. 4 and 5 show us the variation of positive value of
$Im[\omega]$ against $a_{0}$ (keeping $a_{1}$ fixed) and against
$a_{1}$ (keeping $a_{0}$ fixed) respectively. In Fig. 4, the
different values of $a_{1}$ are: 0(blue), .5(red), 1(yellow),
1.5(green) and in Fig. 5, the different values of $a_{0}$ are:
0(blue), .5(red), 1(yellow), 2(green)}}. \vspace{1mm}
\end{figure}

For Figures (2-5) we have taken the positive part of
$\omega$ from the expression (\ref{n3}). Similar analysis could be
done with the negative part. In Figures 2 (and 3) we plot $Re
[\omega] $ vs. $a_0 (a_1)$ keeping $a_1 (a_0)$ fixed. In Figure 2
notice that $a_1=0$ (blue line) is singular at $a_0=1$ (as in
Figure 1) but for non-zero $a_1$ values the lines are not singular
near $a_0=1$. In Figure 3 also $a_0=1$ (yellow line) diverges at
$a_1=0$ since that corresponds to $a_0=1$ from (\ref{n2}).\\
More interesting features are found in Figures 4 (and 5) where we
plot $Im [\omega] $ vs. $a_0 (a_1)$ keeping $a_1 (a_0)$ fixed. We
point out that when $a_0$ and $a_1$ obey the relation (\ref{n2})
the $Im [\omega] $ falls to zero giving real values of energy.
Notice that for complex $a$ the previous singularity (see in
Figures 4 (and 5)) at $a=1$ disappears. Furthermore we show that
it is possible to have real energy for even for $\mid a \mid <1$
whereas for real $a$, real energy is possible only for $a>1$ (see Figure 1). \\
In Figure 4,the expression for $Im [\omega] $ is symmetric in
$a_1$ and for $\mid a_1\mid >1, ~Im [\omega] $ (green line) never
reaches zero.\\
In Figure 5, for $a_0<0$ the lines are entirely in positive $Im
[\omega] $ sides and never vanish as they are outside the bounds
in (\ref{n2}). But for  $a_0>2$,~$Im[\omega] $ will vanish once at
$a_1=0$ which belongs to the normal behavior and so is not shown
in the Figure.\\
We now show that for complex $a$ energy can be real and $\omega<2$
contrary to the case for real $a$ where $\omega \ge 2$. As an
example we put $\omega =1$ in the energy expression (\ref{p1}). It
gives $a=(1\pm i\sqrt 3)/2$ and from Figure 5 we find that for
$a_0=0.5,~~Im[\omega] =0$ (yellow line) at $a_1=\pm \sqrt 3/2$.
Also from Figure 3 for $a_0=0.5$ (red line) we find $Re [\omega]
=1$ for $a_1=\pm \sqrt 3/2$. In fact for $2>a_{0}>0$ all the
states having energy $<2$ are now allowed. \\

\section {\bf{(Quantum) Field Theoretic Model:}}
 Now we come to the parent field theory action in $1+1$-dimensions, that is
 the Chiral Schwinger Model
(CSM) \cite{jr,sch,s1}:
\begin{equation}
S(A , \phi) = \int d^{2}x \left[-\frac{1}{4} F_{\mu \nu} F^{\mu
\nu} + \frac{1}{2}\partial_{\mu}\phi
\partial^{\mu}\phi + e(g^{\mu \nu} - \epsilon^{\mu
\nu})A_{\nu}(\partial_{\mu}\phi) + \frac{1}{2} a e^{2}
A_{\mu}A^{\mu}\right]\label{p3}
\end{equation}
which explicitly turns out to be
\begin{equation}
S = \int d^2x\left[ -\frac{1}{2}F_{0
1}F^{01}+\frac{1}{2}(\partial_{0}\phi)^{2}-\frac{1}{2}
(\partial_{1}\phi)^{2}
+e(\partial_{0}\phi)A_{0}+e\phi(\partial_{0}A_{1})
+e(\partial_{1}\phi)(A_{0}-A_{1})\right.$$$$\left.
+\frac{1}{2}ae^{2}(A_{0}^{2}-A_{1}^{2})\right]\label{p2},
\end{equation}
where in (\ref{p2}) we have dropped a total time derivative term
and in all the further calculations we will use the expression
(\ref{p2}). Notice that in the limit of ignoring the spatial
dependence (thereby dropping space derivatives) (\ref{p2}) reduces
to (\ref{n1}) and (\ref{p5}) (when $a=1$). The parameter $a$,
known as the Jackiw-Rajaraman (JR) parameter appears in the
bosonized model (\ref{p2}) as a result of regularization ambiguity
in evaluating the fermion determinant \cite{s1}. It is taken as a
real number. Once again we will find that since the value of JR
parameter $a$ governs the constraint structure, it can alter the
spectra although no qualitative changes are manifested in the
Lagrangian or Hamiltonian of the model.\\

{$\mathbf{a \neq 1}$}:~~~From the momenta:~~$
\pi_{0}=0$~,~$\pi_{1}=-(\partial_{0}A_{1}-\partial_{1}A_{0})+e\phi=-F_{0
1}$~,~$\pi=\partial_{0}\phi + eA_0$, where~~$\{\phi(x),\pi(y)\} =
\delta(x-y)~,~\{A_{\mu}(x),\pi^{\nu}(y)\} =
g^{\nu}_{\mu}\delta(x-y)$ , we obtain the Hamiltonian:
\begin{equation}
H =\int dx\left(\frac{1}{2}\pi^{2}+\frac{1}{2}\pi_{1}^{2}
+A_{0}(\partial_{1}\pi_{1})+\frac{1}{2}(\partial_{1}\phi)^{2} -e
\pi A_{0}+\frac{1}{2}e^{2}\phi^{2}
+\frac{1}{2}e^{2}A_{0}^{2}\right.$$$$\left.-\frac{1}{2}ae^{2}(A_{0}^{2}-A_{1}^{2})
+e(\partial_{1}\phi)A_{1}+e\phi \pi_{1}\right)
\end{equation}
and two constraints: $ \psi_{1}=\pi_{0}$~,~$\psi_{2}=
\dot{\psi_{1}} = \{\psi_{1},H\}=(a-1)e^{2} A_{0} +e\pi
-\partial_{1}\pi_{1}$~ satisfying the algebra:
$\{\psi_{1}(x),\psi_{2}(y)\}=-(a-1)e^{2}\delta(x-y)$. Again for $a
\ne 1$ the constraints are SCC . We eliminate $A_{0}$ and
$\pi_{0}$ using the constraint equations and for rest of the
variables we see that the Dirac brackets remain same. The reduced
Hamiltonian is:
\begin{equation}
H = \int dx
\left(\frac{1}{2}\pi^{2}+\frac{1}{2}\pi_{1}^{2}+\frac{1}{2}e^{2}\phi^{2}
+\frac{1}{2}(\partial_{1}\phi)^{2}
+e(\partial_{1}\phi)A_{1}+e\phi
 \pi_{1}\right.$$$$\left.+\frac{1}{2}ae^{2}A_{1}^{2}
 +\frac{1}{2e^{2}(a-1)}\left((\partial_{1}\pi_{1})^{2}+e^{2}\pi^{2}
-2e\pi(\partial_{1}\pi_{1})\right)\right)
\end{equation}
which yields the spectra, consisting of a Klein-Gordon scalar
$\sigma = \pi_{1}+e\phi $ and a harmonic mode $h =
(\pi-eA_{1})-\frac{1}{ae}\partial_{1}(\pi_{1}+e\phi)$,
\begin{equation}
\Box \sigma + m^{2} \sigma = 0~~;~~\Box h = 0~~;~~m^{2} =
\frac{a^{2} e^{2}}{a-1}. \label{ex}
\end{equation}
The theory is consistent for $a>1$ otherwise there are tachyonic
excitations. Notice that $h$ satisfies a higher derivative equation. These modes
reduce to the previously computed spectra
(\ref{p1}) when the space dependence is ignored.\\

{$\mathbf{a=1}$}:~~~We directly put $a=1$ in (\ref{p2}),
\begin{equation}
S = \int d^2x\left[ -\frac{1}{2}F_{0
1}F^{01}+\frac{1}{2}(\partial_{0}\phi)^{2}-\frac{1}{2}
(\partial_{1}\phi)^{2}
+e(\partial_{0}\phi)A_{0}+e\phi(\partial_{0}A_{1})
+e(\partial_{1}\phi)(A_{0}-A_{1})\right.$$$$\left.
+\frac{1}{2}e^{2}(A_{0}^{2}-A_{1}^{2})\right]\label{p20},
\end{equation}
and obtain the Hamiltonian
\begin{equation}
H =\int dx \left(
\frac{1}{2}\pi^{2}+\frac{1}{2}\pi_{1}^{2}+\frac{1}{2}e^{2}A_{1}^{2}
+\frac{1}{2}e^{2}\phi^{2}
+\frac{1}{2}(\partial_{1}\phi)^{2}-e\pi
A_{0}+A_{0}(\partial_{1}\pi_{1})+e(\partial_{1}\phi)A_{1}+e \phi
\pi_{1}\right).
\end{equation}
Now there are three constraints,
$$\psi_{1} \equiv \pi_{0},~~~\psi_{2} \equiv
\dot{\psi_{1}}=\{\psi_{1},H\}=e\pi-(\partial_{1}\pi_{1}),$$
\begin{equation}
\psi_{3} \equiv \dot{\psi_{2}}=\{\dot{\psi_{2}},H
\}=\pi_{1}+e\phi. \label{a}
\end{equation}
Using Dirac's procedure \cite{dir} we find that $\psi_1$ is an FCC
and $\psi_2,\psi_3$ constitute an SCC pair. Here the canonical
bracket between the variables $(\phi,\pi)$ remains unchanged.
Using the constraint equations we have the Hamiltonian and a
single massless mode $h=(\partial_{0}^{2}+\partial_{1}^{2})(\phi)$
respectively
\begin{equation}
H=\int dx
\left[\frac{1}{2}(\partial_{1}\phi)^{2}+\frac{1}{2}(\partial_{1}\phi
+eA_{1})^{2}\right]~~~,~~~
\Box h=0 \label{h}
\end{equation}

which agrees with the spectrum in the particle limit. This shows
that the characteristic features associated with $a=1$ in the
previously analyzed model remain intact in the field theory as
well.

Note that in both $a=1$ and $a\neq 1$ cases the {\it{massless}}
modes satisfies higher derivative equations. In particular $h$ in
(\ref{h}) satisfy a fourth order equation that is clearly
reminiscent of the mode dynamics in Pais-Uhlenbeck oscillator
\cite{pais} that we have advertised in the Introduction.

In passing we make a generic comment. Notice that presence of
gauge invariance fixes the value of $a$ to $a=1$. An analogous
situation prevails in the bosonization of the vector Schwinger
model \cite{jr,sch} which has gauge symmetry and expectedly no
arbitrary parameter appears in its bosonized version.\\

\section {\bf{Cranking (particle) Model and
Maxwell-Chern-Simons-Proca (field) Theory: }}
 We briefly mention the connection between the Cranking Model \cite{val} which is well
known in Nuclear Physics and also recently studied \cite{h2} in
the context of EP and its close connection with a widely studied
relativistic field theory in $2+1$-dimension; the
Maxwell-Chern-Simons-Proca (MCSP) model \cite{djt,rb,sg}. We show
that the connection is similar as the model (\ref{p1}) and CSM
(\ref{p2}) considered in Sections 3 and 4 if we drop the spatial derivatives.
We consider the two-particle Lagrangian \cite{rb,sg}:
\begin{equation}
L=\frac{1}{2}\dot{x_{1}}^{2}+\frac{1}{2}\dot{x_{2}}^{2}
+\frac{B}{2}(x_{1}\dot{x_{2}}-x_{2}\dot{x_{1}})
-\frac{k}{2}(x_{1}^{2}+x_{2}^{2}).
\end{equation}
Using the momenta
$p_{i}=\dot{x_{i}}-\frac{B}{2}\epsilon_{ij}x_{j}$ we find the
Cranking Model Hamiltonian \cite{h2},
\begin{equation}
H=\frac{1}{2}[p_{1}^{2}+p_{2}^{2} +(\frac{B^{2}}{4}+k)(x_{1}^{2}
+x_{2}^{2}) -B(x_{1}p_{2}-x_{2}p_{1})].
\end{equation}
We have scaled the masses to unity. After Bogoliubov
transformation the above Hamiltonian becomes diagonal where the
energy eigenmodes are given by:
\begin{equation}
\omega_{\pm}^{2}=\frac{1}{2}(2k+B^2)
\left[1\pm\left(1-\frac{4k^2}{(2k+B^{2})^{2}}\right)
^{\frac{1}{2}}\right].\label{p3}
\end{equation}

We emphasize that this model can be obtained from MCSP model,
\begin{equation}
L_{MCSP}=-\frac{1}{4}A_{\mu\nu}A^{\mu\nu}+
\frac{B}{2}\epsilon_{\mu\nu\lambda}
(\partial^{\mu}A^{\nu})A^{\lambda}+\frac{k}{2}A_{\mu}A^{\mu},
\label{p4}
\end{equation}
by ignoring the space dependence and dropping the decoupled $A_0$
term. Taking account of the constraints in (\ref{p4}) and using
the Dirac prescription \cite{dir} we find
\begin{equation}
H=\frac{1}{2}\pi_{i}^{2}
+\frac{1}{4}(A_{ij})^{2}+\left(\frac{1}{2}+\frac{B^{2}}{8}\right)
A_{i}^{2}
-\frac{B}{2}\epsilon_{ij}\pi_{i}A_{j}+\frac{1}{2}
\left(\partial_{i}\pi_{i}+\frac{B}{2}\epsilon_{ij}(\partial_{i}
A_{j})\right)^{2}.
\end{equation}
Applying the following nonlocal canonical transformations \cite{djt,rb,sg}
\begin{equation}
A_{i}=\epsilon_{ij}
\frac{\partial_{j}(Q_{1}+Q_{2})}{\sqrt{-\nabla^{2}}}+\frac{1}{2}
\frac{\partial_{i}(P_{1}-P_{2})}{\sqrt{-\nabla^{2}}}~~;~~\pi_{i}=
\frac{1}{2}\epsilon_{ij}
\frac{\partial_{j}(P_{1}+P_{2})}{\sqrt{-\nabla^{2}}}
-\frac{\partial_{i}(Q_{1}-Q_{2})}{\sqrt{-\nabla^{2}}}
\end{equation}
the Hamiltonian becomes decoupled in the form:
\begin{equation}
H=\left[\frac{1}{2}(P_{1}^{2}+(\partial_{i}Q_{1})^{2}
+M_{1}^{2}Q_{1}^{2})
+\frac{1}{2}(P_{2}^{2}+(\partial_{i}Q_{2})^{2}
+M_{2}^{2}Q_{2}^{2})\right].
\end{equation}
Now $M_1,M_2$ are identical to $\omega_{\pm}$ in (\ref{p3}) and
again ignoring spatial derivatives the above becomes identical to
the spectra (\ref{p3}). This is our advertised correspondence.

\section{Conclusion and Future Prospects}
In the present paper we have demonstrated how specific value of a
single parameter (in the model the JR parameter $a$) can influence
the entire dynamical content of a model. This point in parameter
space is exceptional in the sense that the nature of the
constraint structure is dictated by this specific point , i.e. at
$a=1$ the system has First Class Constraints, inducing local gauge
invariance with a reduced number of physical degrees of freedom
whereas away from $a=1$ the system has only Second Class
Constraints with no additional invariance and so possessing a
larger number of degrees of freedom. These results, in the field
theoretic Chiral Schwinger model are not new but the results in
the corresponding finite dimensional particle model that we have
formulated and studied are indeed new and interesting. We have
explored the complex domain of the JR parameter $a$ that reveals
the existence of real and physically realizable energy values that
were forbidden in the field theory context. We have also revealed
the connection between two  well known discrete and field
theoretic models, the Cranking model and the
Maxwell-Chern-Simons-Proca model, studied in entirely different
contexts. These works actually lead us to two interesting and
topical areas, mentioned below, that we wish to pursue.

First of all note that just now we have referred to the $a=1$
point in the parameter point as exceptional for a specific reason.
In fact we believe that this point might be an interesting example
of a novel type of Exceptional Point \cite{he1,h2,kato,others} in the context
of non-Hermitian $PT$-symmetric quantum mechanics.
The point $a=1$ shares several features of the conventional
Exceptional Point in the sense that $a=1$ point lies in the border
of real and imaginary energy values (or non-unitary spectra in
case of field theory); there is an (apparent) singularity as
$a\rightarrow 1$ (although the $a=1$ point is not singular as
further analysis shows); there is a collapse of states (in this
case an infinite tower of states). (For qualitatively similar
effects in the context of Exceptional Points see \cite{he1,h2}.)
Furthermore the Cranking model has already been studied in the
context of Exceptional Points and $PT$-symmetric quantum mechanics
and we have discussed here similar behavior for the discrete
version of chiral Schwinger model having real energy values for
complex parameter $a$ present in the Hamiltonian. These results
can pave the way for  study of these features {\it{i.e.}}
$PT$-symmetry and Exceptional Points in the corresponding
relativistic field theory models such as chiral Schwinger model
and Maxwell-Chern-Simons-Proca model.


\begin{thebibliography}{99}
\bibitem{s1}The CSM was derived from a theory of chiral fermions
interacting with electromagnetic field in $1+1$-dimensions $ S =
\int d^{2}x \left[-\frac{1}{4} F_{\mu \nu} F^{\mu \nu} +
\bar{\psi}[i \mathbf{\partial} + e \sqrt{\pi} \mathbf{A}(1 + i
\gamma_{5})] \psi\right] $ where $ \gamma_{5} = i \gamma^{0}
\gamma^{1} $ via bosonization.
\end{thebibliography}
\end{document}